\documentclass[preprint,12pt,english]{revtex4-1}
\usepackage[latin9]{inputenc}
\usepackage[fleqn]{amsmath}
\usepackage{mathptmx}
\usepackage{amssymb}
\usepackage{mathtools}
\usepackage{url}
\usepackage{graphicx}
\usepackage{caption}
\usepackage{subcaption}
\usepackage{epstopdf}
\usepackage{esint}
\usepackage{lineno,hyperref}
\modulolinenumbers[5]
\usepackage{babel}
\usepackage{color}
\usepackage{ulem}

\DeclareSymbolFont{matha}{OML}{txmi}{m}{it}
\DeclareMathSymbol{\vrv}{\mathord}{matha}{118}
\makeatother
\usepackage{babel}

\begin{document}
\title{Stochastic resetting and first arrival subjected to Gaussian noise and Poisson white noise}
\author{Koushik Goswami and Rajarshi Chakrabarti*}
\address{Department of Chemistry, Indian Institute of Technology Bombay, Mumbai, Powai 400076, India. *E-mail: rajarshi@chem.iitb.ac.in}
 \author{\textbf{Published in Phys. Rev. E 104, 034113 (2021) }}

\begin{abstract}
\noindent We study the dynamics of an overdamped Brownian particle subjected to Poissonian stochastic resetting in a nonthermal bath, characterized by a Poisson white noise and a Gaussian noise. Applying the renewal theory we find an exact analytical expression for the spatial distribution at the steady state. Unlike the single exponential
distribution as observed in the case of a purely thermal bath, the distribution is double exponential. Relaxation of the transient spatial distributions to the stationary one, for the limiting cases of Poissonian rate, is investigated carefully. In addition, we study the first-arrival properties of the system in the presence of a delta-function sink
with strength $\kappa$, where $\kappa=0$ and $\kappa=\infty$ correspond to fully nonreactive and fully reactive sinks, respectively.
We explore the effect of two competitive mechanisms: the diffusive spread in the presence of two noises and the increase in probability density around the initial position due to stochastic resetting. We show that there exists
an optimal resetting rate, which minimizes the mean first-arrival time (MFAT) to the sink for a given value of the sink strength. We also explore the effect of the strength of the Poissonian noise on MFAT, in addition to sink strength. Our formalism generalizes the diffusion-limited reaction under resetting in a nonequilibrium bath
and provides an efficient search strategy for a reactant to find a target site, relevant in a range of biophysical
processes
\end{abstract}

\maketitle
\section{Introduction \label{intro}}

\noindent  In the mesoscopic world,  diffusion refers to the random motion of Brownian particles in a fluid at thermal equilibrium.  In this case, the detailed balance as well as congruent features such as the  fluctuation-dissipation theorem (FDT) and zero heat flux are maintained \cite{einstein1905movement}. In other words, diffusion is a passive process as it takes place in thermal equilibrium.  In recent years, there has been growing interest in studies of active particles, which are self-driven due to the consumption of energy from their surroundings and therefore away from equilibrium \cite{bechinger2016active,Caprini_2021,grandpre2021entropy}. 
Models such as active Brownian particles (ABPs) and run-and-tumble particles (RTPs) have been used extensively for their dynamical descriptions incorporating the  persistence in their motion  \cite{ramaswamy2017active,fodor2018statistical,martin2021statistical}. Evidently, when a passive   molecule such as colloid or polymer is immersed in a bath containing active particles, the molecule  behaves differently from a thermal (or equilibrium) bath \cite{goswami2019diffusion,goswami2021nonequilibrium,chaki2018entropy}. Apart from the thermal noise, it experiences  additional nonequilibrium fluctuations (active noise) which drives the system away from equilibrium \cite{samanta2016chain,physRevLett.116.248301,gladrow2017nonequilibrium,chaki2019effects,chaki2018entropy,goswami2019heat,goswami2021work,goswami2019work}. The peculiarities of its dynamics are often encoded in its distribution function which, in some cases, is  found to be non-Gaussian in nature \cite{toyota2011non,PhysRevE.86.020901,fodor2015activity}. For instance, the motion of a tracer particle inside the cytoskeleton network  has been found to be non-Gaussian consisting of two parts: the Gauss-like central region followed by an exponential tail \cite{toyota2011non}. Such tail occurs due to the athermal noise stemming from the power stroke generated by motor proteins, and the noise can be conceived as the Poisson one. In practice, the Poissonian noise has been widely used  as an effective model for non-Gaussianity in biological systems, for instances, to explain membrane undulations \cite{gov2004membrane,ben2011effective}, neuron dynamics \cite{PhysRevLett.92.080601}, swelling of a polymer \cite{chaki2019enhanced}, etc.  If the system does not have any memory effect, or roughly speaking, the dynamics is Fickian, not necessarily, one should invoke a finite correlation time in the noise statistics to describe it . Sometimes memoryless fluctuations can render an out-of-equilibrium state  \cite{PhysRevLett.108.210601,PhysRevLett.114.090601,PhysRevE.93.012121,fodor2018non,goswami2019heat,goswami2019work}. For this matter, the Poissonian shot noise is an ideal choice for characterization of an athermal bath to study Markovian, non-Gaussian dynamics \cite{goswami2021nonequilibrium,tgera2021solution}. Note that such noise has been previously employed for modeling several physical cases such as ATP hydrolysis in the context of movement of motor proteins \cite{reimann2002brownian}, nonperiodic oscillatory distortion of a lipid interface coupled with actomyosin \cite{nishigami2016non}, and dynamics of a granular rotor in an environment containing low-density granular gases \cite{PhysRevLett.114.090601,sano2016granular}. 

 Search processes are very much intertwined with diffusion for small systems, particularly ones involved in diffusion-limited reactions. For example, during gene expression, a protein molecule (transcription factor) searches for a promoter site on the DNA chain to initiate transcription. Along with others, two main routes it adopts to find the target are three-dimensional excursions in the surroundings and one-dimensional diffusive search (sliding) along the DNA track \cite{berg1985diffusion, metzler2012biophysj, bagchi2009NCSB}. A major goal in investigating such processes is to find a strategy for which the rate gets minimized.  Along these lines, a lot of attention has rightfully been paid these days to the stochastic resetting, a mechanism for which a  system undergoing a stochastic process is reset back at random times to a prescribed position and restarts the process all over again \cite{PhysRevLett.106.160601}. Two key features of diffusion processes with resetting are that (i) the system always achieves  a nontrivial nonequilibrium steady state, and  (ii) in the presence of a target, there exists an optimal rate for which the search time is finite and minimum \cite{evans2013optimal,evans2020stochastic}. No wonder the occurrence of such intriguing characteristics triggers exploring its different variants, namely, resetting with several processes such as continuous-time random walks \cite{mendez2021continuous}, fractional Brownian motions \cite{majumdar2018spectral}, L\'evy flights \cite{PhysRevE.92.052127}, underdamped diffusion \cite{gupta2019stochastic},   velocity-jump processes \cite{bressloff2020directed} and others \cite{PhysRevResearch.2.033182,basu2019symmetric}. Furthermore, the effect of various confining potentials \cite{PhysRevE.96.022130,singh2020resetting,mercado2020intermittent} and diffusivities \cite{bressloff2020switching} on the resetting mechanism has been investigated thoroughly over the past few years. 

The majority of the studies on resetting mentioned above have been carried out for diffusive systems in the thermal bath. Recently, a number of groups have explored the possibility of incorporating the activity in the resetting mechanism \cite{mori2020universal,masoliver2019telegraphic,santra2020run,evans2018run}. A charged ABP subjected to a spatially nonhomogeneous magnetic field has been found to have a nonmonotonic density profile and a higher mean first-passage time compared to its passive counterpart \cite{abdoli2021stochastic}. Further developments have witnessed implications of stochastic resetting  in the dynamics of RTPs which serves as the  standard  model for bacterial motion \cite{masoliver2019telegraphic,evans2018run}. However, to the best of our knowledge, very little is known about the role of nonthermal fluctuations on a passive particle under a resetting mechanism. In this paper, our interest is to fill  the existing gap in the literature, and try to provide a comprehensive study on the dynamics and first-passage properties of an activity-driven stochastic resetting process. To do so, here we consider a situation where a passive Brownian particle is diffusing in a non-thermal bath characterized by the Poissonian white noise. The particle is brought back randomly at a constant rate $r$ to its initial position $x_0,$ from where it restarts its journey following the underlying stochastic equation.  Note that in our model,  the epochs at which the particle returns to $x_0$ are drawn from a Poissonian distribution; that is to say, the waiting time distribution is given by $\psi_w(t)=r e^{- r t}.$  However,   it may be possible to generalize the resetting  mechanism by considering a generic form of $\psi_w(t);$  $e.g.,$ see Ref. \cite{masoliver2019anomalous}. In Sec. \ref{model} we  illustrate  the effect of resetting on  dynamical behaviors of the particle diffusing in an unbounded domain.  At large times  the particle relaxes to a stationary state, and the relaxation mechanisms for two limiting cases are discussed in Sec. \ref{relaxation}. Further we investigate the first-arrival properties of  the system in the presence of a sink (or target) at some position $x_S$. When the particle meets the target, either it binds to the target instantaneously, $i.e.,$ it gets fully absorbed at $x=x_S$, or it may not identify the target with certainty which is referred to as the partial absorption \cite{whitehouse2013effect}. One common interest in such study is to find the rate and efficiency of the process.  The efficiency is generally determined by the time  it  takes on average to reach the sink for the first time, which is called the mean first-arrival time (MFAT), and the distribution of  MFAT known as the first-arrival time density (FATD) is a measure of the  instantaneous rate. By finding those quantities, we demonstrate how  first-arrival properties depend on the interplay between activity and resetting mechanism, as discussed in Sec. \ref{First-passage-properties}. It should  be noted here that our theoretical model can be realized experimentally by mimicking the recent experimentation on the stochastic resetting \cite{tal2020experimental}  to corroborate the results. Finally, the summary of our findings is given in Sec. \ref{conclusion}.

\section{ Stochastic resetting in a free space \label{model}}

\noindent Suppose a Brownian particle is  moving in a free space in the  nonthermal bath. It is subjected to the thermal noise $\eta_T(t)$ and an athermal noise $\xi_A(t).$  $\eta_T(t)$ is usually given by  the symmetric, delta-correlated Gaussian noise  which obeys the fluctuation-dissipation theorem (FDT): $\langle \eta_T(t_1) \eta_T(t_2)\rangle=2 D_T \delta(t_1-t_2),$ where $D_T$  is  the diffusivity of the particle in the thermal bath. On the other hand, the noise $\xi_A(t)$ violates the FDT and so  the bath can no longer be described by the equilibrium properties. Here we do not take into account the memory effect, and so  $\xi_A(t)$ is taken as  the  Poissonian white noise. It can be realized as a sequence of delta pulses occurring at random times over a time interval $[0,t],$  $viz.,$
$\xi_A(t)=\sum_{i} a_i \delta(t-t_i) .$  
The occurrence of pulses follows the Poisson statistics with a Poisson rate $\mu,$ and the jump-length associated with $i$th pulse denoted by $a_i$ is drawn from the Laplace distribution of the form $\mathcal{P}(a)=\frac{1}{2 a_0}e^{-\frac{|a|}{a_0}},$ where $a_0$ is the average jump length. It is easy to see that  $\langle \xi_A(t)\rangle =0,$ and $\langle \xi_A(t) \xi_A(t')\rangle =2D_A \delta(t-t'),$   where the diffusivity due to the non-thermal noise is given by $D_A=\mu a_0^2$ \cite{goswami2021nonequilibrium}.  

Let us consider that the particle is initially at  position $x_0,$ and the probability density at position $x$ after time $t$ is $P_0(x,t).$  The density, or the propagator $P_0(x,t|x_0)$ evolves according to  the following  Kolmogorov-Feller equation \cite{dubkov2016probability,fodor2018non}: 
\begin{align}
    \frac{\partial }{\partial t}P_0(x,t)=D_T \frac{\partial^2  }{\partial x^2}P_0(x,t)+ D_A \frac{\frac{\partial^2}{\partial x^2}}{1-a_0^2 \frac{\partial^2}{\partial x^2}}P_0(x,t)\label{FP0}.
\end{align}
For an alternative derivation of the above equation, see Appendix  \ref{Appen1}.  
 In Eq. (\ref{FP0}), the first term on the right-hand side (RHS) is given by a second-order derivative  which describes  the normal diffusion.  In contrast to this, the final term contains a nonlocal differential operator which corresponds to the Poissonian noise. If one consider the limit $a_0 \rightarrow 0$ and $\mu  \rightarrow \infty$ in such a way that $D_A$ remains constant, the nonlocal property vanishes so that one  recovers the normal diffusion case \cite{van1983relation}. In general, an exact analytical solution of Eq. (\ref{FP0}) is difficult to obtain. However, it can be treated analytically using the Fourier transform, defined as $\Tilde{P}_0(p,t)=\frac{1}{2\pi}\int dx\,e^{-i p x}P_0(x,t).$ On doing the transform, Eq. (\ref{FP0}) becomes 
 \begin{align}
\Tilde{P}_0(p,t)=e^{-D_T p^2 t-D_A t \frac{p^2}{1+a_0^2 p^2}}\label{FP0_fourier}.
 \end{align}
 With this result, one can find  several limiting cases for the spatial density. For more details, the reader is referred to Ref. \cite{goswami2019diffusion}.
 
\noindent We now address the resetting problem for the above stochastic process. Here  we consider the  Poissonian resetting protocol; $i.e.,$ the waiting times between two successive resetting events follow the exponential distribution with rate $r.$ Also  let us assume that the particle is reset to a resetting position which, for simplicity, is the same as the initial position $x_0$. For this case, the probability distribution function (PDF), denoted by $P_r(x,t),$ satisfies the equation
\begin{align}
    \frac{\partial }{\partial t}P_r(x,t)=D_T \frac{\partial^2  }{\partial x^2}P_r(x,t)+ D_A \frac{\frac{\partial^2}{\partial x^2}}{1-a_0^2 \frac{\partial^2}{\partial x^2}}P_r(x,t) - r\,P_r(x,t)+r \delta(x-x_0) \label{FP_resetting}.
\end{align}
The last two terms account for the loss and gain of probabilities due to resetting, respectively.
Taking the Fourier transform of  Eq. (\ref{FP_resetting}), one  obtains 
\begin{align}
   \frac{\partial }{\partial t}\Tilde{P}_r(p,t)= -D_T p^2  \Tilde{P}_r(p,t)-D_A \frac{p^2}{1+a_0^2 p^2}\Tilde{P}_r(p,t)-r \Tilde{P}_r(p,t) +r e^{-i p x_0}.\label{renewal_Fourier0}
\end{align}

Without loss of any generality, here we take  $x_0=0,$ which makes the initial condition to be $P_r(x,0)=\delta(x).$ After  performing simple mathematical steps and rearranging the terms as shown in Appendix \ref{appen2},  we obtain
\begin{equation}
    \Tilde{P}_r(p,t)=e^{-r t} \Tilde{P}_0(p,t)+r\int_{0}^{t}\, dt'\,e^{-r t'} \Tilde{P}_0(p,t')\label{renewal_Fourier}.
\end{equation}
Upon doing the inverse Fourier transformation of the above, we have 
\begin{align}
P_r(x,t)=e^{-r t} P_0(x,t)+r\int_{0}^{t}\, dt'\,e^{-r t'} P_0(x,t')\label{renewal_eqn},
\end{align}
where $P_0(x,t)$ is the solution of Eq. (\ref{FP0}), and it can be expressed as the inverse Fourier transform of Eq. (\ref{FP0_fourier}), $i.e.$,
$P_0(x,t)=\int \frac{dp}{2\pi} \,e^{-i p x}\,e^{-D_T p^2 t-D_A t \frac{p^2}{1+a_0^2 p^2}}.
$ 
 One can easily recognize Eq. (\ref{renewal_eqn}) as the last  renewal equation for this process.

Now our motive is to obtain the density $P_r(x,t).$ For that, we go back to Eq. (\ref{renewal_Fourier}), and writing the expression for $\Tilde{P}_0(p,t)$  explicitly  with the aid of Eq. (\ref{FP0_fourier}), we have  
\begin{align}
  \Tilde{P}_r(p,t)=e^{-r t}\,e^{-\left(D_T t p^2 +D_A  t \frac{p^2}{1+a_0^2 p^2} \right)}+r\int_{0}^{t}\, dt'\,e^{-r t'}\,e^{-\left(D_T t' p^2 +D_A  t' \frac{p^2}{1+a_0^2 p^2} \right)} \label{renewal_Fourier1}.
\end{align}

 In the final term, the integration over time can be easily done, and it yields 
 
\begin{align}
\Tilde{P}_r(p,t)=e^{-r t}\,e^{-\left(D_T t p^2 +D_A  t \frac{p^2}{1+a_0^2 p^2} \right)} +\frac{r(1+a_0^2p^2)\left[1-e^{-r t}\,e^{-\left(D_T t p^2 +D_A  t \frac{p^2}{1+a_0^2 p^2} \right)}\right]}{D_T a_0^2 p^4+(D_A +D_T+a_0^2 r)p^2+ r}\label{renewal_Fourier1}.
\end{align}

Further rearranging the terms,  we arrive at an analytically tractable form as 
\begin{align}
 \Tilde{P}_r(p,t)=& e^{-r t}\,e^{-\left(D_T t p^2 +D_A  t \frac{p^2}{1+a_0^2 p^2} \right)}+\frac{r}{ D_T} \left[1-e^{-r t}\,e^{-\left(D_T t p^2 +D_A  t \frac{p^2}{1+a_0^2 p^2} \right)}\right] \nonumber\\
 & \quad\quad \quad\quad\quad\quad\times \left[\left(\frac{\alpha_3(r)-\alpha_1(r)}{\alpha_2(r)-\alpha_1(r)}\right) \frac{1}{p^2+\alpha_1(r)}+\left(\frac{\alpha_2(r)-\alpha_3(r)}{\alpha_2(r)-\alpha_1(r)}\right) \frac{1}{p^2+\alpha_2(r)}\right]\label{renewal_Fourier2},
\end{align}
where 
$\alpha_3(r)\equiv \alpha_3=\frac{1}{a_0^2},\,\alpha_2(r)+\alpha_1(r)=\frac{D_A}{D_T}\frac{1}{a_0^2} +\frac{1}{a_0^2}+\frac{r}{D_T},\,\alpha_1(r)\alpha_2(r)= \frac{r}{D_T}\frac{1}{a_0^2},$ which implies

$$\alpha_2(r)=\frac{1}{2a_0^2}\left[\frac{D_A}{D_T} +1+\frac{a_0^2 r}{D_T}\right]+\frac{1}{2a_0^2}\sqrt{\left[\frac{D_A}{D_T} +1+\frac{a_0^2 r}{D_T}\right]^2- \frac{4 a_0^2 r}{D_T } },$$

$$\alpha_1(r)=\frac{1}{2a_0^2}\left[\frac{D_A}{D_T} +1+\frac{a_0^2 r}{D_T}\right]-\frac{1}{2a_0^2}\sqrt{\left[\frac{D_A}{D_T} +1+\frac{a_0^2 r}{D_T}\right]^2- \frac{4 a_0^2 r}{D_T } }.$$

Now we can perform  the inverse Fourier transform  of Eq. (\ref{renewal_Fourier2}), and it reads
\begin{align}
 & P_r(x,t)= e^{-r t}\,P_0(x,t)+\frac{r}{D_T} \left[\left(\frac{\alpha_3(r)-\alpha_1(r)}{\alpha_2(r)-\alpha_1(r)}\right) \frac{e^{-\sqrt{\alpha_1(r)} |x|}}{2\sqrt{\alpha_1(r)}}+\left(\frac{\alpha_2(r)-\alpha_3(r)}{\alpha_2(r)-\alpha_1(r)}\right) \frac{e^{-\sqrt{\alpha_2(r)} |x|}}{2\sqrt{\alpha_2(r)}}\right]\nonumber\\
&-\frac{r\,e^{-r t}}{D_T}\int\,dx'\,\left[\left(\frac{\alpha_3(r)-\alpha_1(r)}{\alpha_2(r)-\alpha_1(r)}\right) \frac{e^{-\sqrt{\alpha_1(r)} |x'|}}{2\sqrt{\alpha_1(r)}}+\left(\frac{\alpha_2(r)-\alpha_3(r)}{\alpha_2(r)-\alpha_1(r)}\right) \frac{e^{-\sqrt{\alpha_2(r)} |x'|}}{2\sqrt{\alpha_2(r)}}\right]\,P_0(x-x',t).\label{P_r_final}
\end{align}

The moment of the displacement  can be calculated from the Fourier transform (\ref{renewal_Fourier2}), using the relation, $\langle x^n(t)\rangle=(-i)^n\frac{\partial^n \Tilde{P}_r(p,t)}{\partial p^n}\Big|_{p=0}.$ Since the process is symmetric in space, all odd moments are zero. The second moment or the mean square displacement (MSD) is  computed  as 
\begin{align}
\langle x^2(t)\rangle=\frac{2}{r}(D_A+D_T)\left(1-e^{-r t}\right).
\end{align}
For short-time limits, $i.e.,$ for $t\ll 1/r,$ the MSD is $\langle x^2(t)\rangle \approx 2(D_A+D_T) t,$   implying a short-time diffusive regime, and it is the same as the reset-free case. However, in long-time limits $(t\gg 1/r)$, the MSD approaches a constant value, $viz.,$ $\langle x^2(t)\rangle \approx \frac{2(D_A+D_T)}{r},$ which indicates the convergence of density to a stationary distribution.

Here we analyze the distribution  for  the  $t\rightarrow \infty$ limit. In this case, the first and third terms on  the RHS of  Eq. (\ref{P_r_final}) can be neglected, and so the distribution becomes
\begin{align}
P_{r,ss}(x)=\frac{r}{D_T}\left[\left(\frac{\alpha_3(r)-\alpha_1(r)}{\alpha_2(r)-\alpha_1(r)}\right) \frac{e^{-\sqrt{\alpha_1(r)} |x|}}{2\sqrt{\alpha_1(r)}}+\left(\frac{\alpha_2(r)-\alpha_3(r)}{\alpha_2(r)-\alpha_1(r)}\right) \frac{e^{-\sqrt{\alpha_2(r)} |x|}}{2\sqrt{\alpha_2(r)}}\right]\label{P_r_ss}.
\end{align}
Clearly, Eq. (\ref{P_r_ss}) corresponds to the steady-state distribution which is expressed as the sum of two exponential functions. To understand its leading behavior, we consider different limiting cases. 

\subsection{Different $r-$limits \label{r limit}}
\begin{figure}
\centering
\begin{subfigure}{.5\textwidth}
 \centering
 \includegraphics[width=1.0\linewidth]{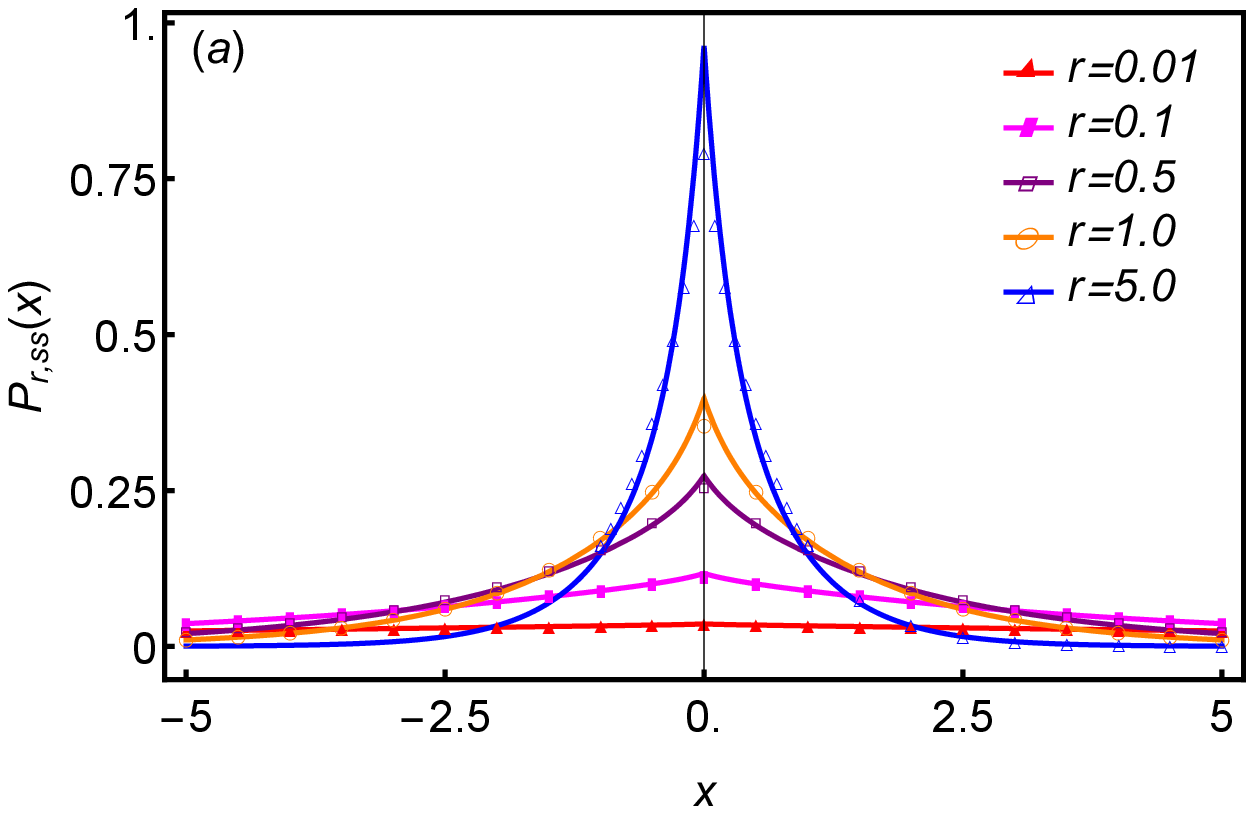}
\end{subfigure}%
\begin{subfigure}{.49\textwidth}
  \centering
 \includegraphics[width=1.0\linewidth]{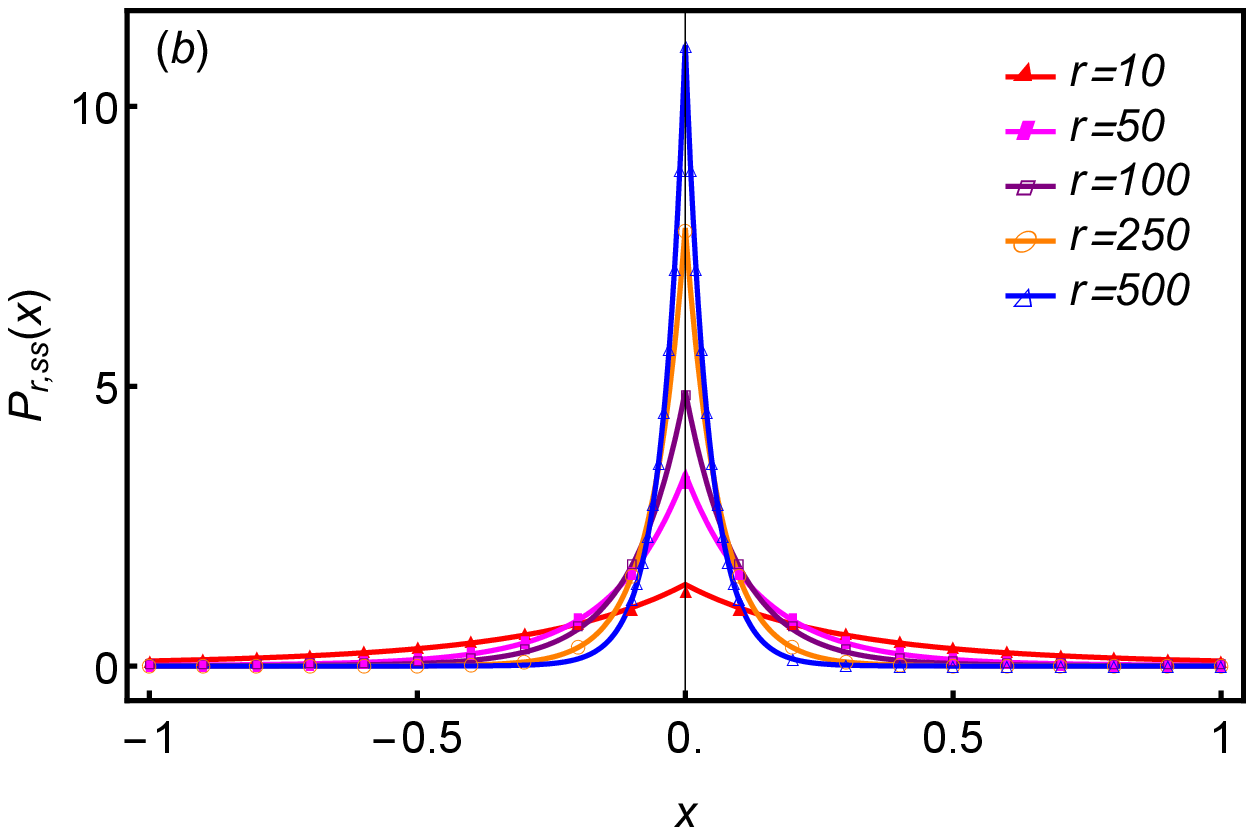}
\end{subfigure}
\caption{Plot of the probability distribution function of displacement at the steady state [given in Eq. (\ref{P_r_ss})] for two different regimes  of Poisson rates:  (a) $r<\mu,$ where $\mu=10,$  and (b)  $r>\mu,$ where $\mu=5.0.$ Here we have taken $D_A=D_T=1.0.$ The  curves with point symbols are the approximate analytical results given in Eqs. (\ref{graph1a}) and (\ref{graph1b}), and these equations are used for comparison in panels (a) and (b),  respectively. }
    \label{plot_pdfshotr}
\end{figure}

Let us first consider the limit $\frac{a_0^2r}{D_T}\left(=\frac{D_A}{D_T}\frac{r}{\mu}\right)\ll 1.$  So using the  approximation $\sqrt{\left[\frac{D_A}{D_T} +1+\frac{a_0^2 r}{D_T}\right]^2- \frac{4 a_0^2 r}{D_T } } \approx \frac{D_A}{D_T} +1-\frac{2a_0^2 r}{D_A+D_T}$, the parameters can be expressed as 
$\alpha_1\approx  \frac{ r}{\left[D_A +D_T\right]}$, $\alpha_2\approx \frac{1}{a_0^2}\left( \frac{D_A}{D_T} +1\right),$ which leads Eq. (\ref{P_r_ss}) to 
\begin{align}
 P_{r,ss}(x) & \approx \frac{1}{2a_0}\left(1-\frac{\left(\frac{a_0^2 r}{D_T}\right)\left(\frac{D_A}{D_T}\right)}{\left(\frac{D_A}{D_T}+1\right)^2}\right) \sqrt{\frac{a_0^2 r}{D_A+D_T}}e^{-\sqrt{\frac{a_0^2r}{D_A+D_T}}\frac{|x|}{a_0}} \nonumber\\
 &+ \frac{1}{2 a_0}\frac{\left(\frac{a_0^2 r}{D_T}\right)\left(\frac{D_A}{D_T}\right)}{\left(\frac{D_A}{D_T}+1\right)^2}\,\sqrt{\left(\frac{D_A}{D_T}+1\right)}\, e^{-\sqrt{\frac{D_A}{D_T}+1}\frac{|x|}{a_0}}.
\end{align}
In the above equation, the second term on the RHS captures the behavior near $x=0,$ and it can be neglected for $\frac{a_0^2r}{D_T}\ll 1.$ Here,  the first term  dominates, and thus the distribution can be approximated as 
\begin{align}
    P_{r,ss}(x) \approx \frac{1}{2} \sqrt{\frac{ r}{D_A+D_T}}e^{-\sqrt{\frac{r}{D_A+D_T}}|x|}.\label{graph1a}
\end{align}
Such limiting case can be obtained if $r\ll \mu$ and $D_A/D_T$ is finite. 
 At $r\rightarrow 0,$ the distribution almost flattens as exemplified in Fig.  \ref{plot_pdfshotr}(a) for $r=0.01.$ Within this regime, if $r> \mu$, then $D_T\gg D_A,$ and thus, $P_{r,ss}(x) \approx \frac{1}{2} \sqrt{\frac{ r}{D_T}}e^{-\sqrt{\frac{r}{D_T}}|x|}.$  

Now we consider the limit  $\frac{a_0^2r}{D_T} \left(=\frac{D_A}{D_T}\frac{r}{\mu}\right)\gg 1.$   In this limit,  $\sqrt{\left[\frac{D_A}{D_T} +1+\frac{a_0^2 r}{D_T}\right]^2- \frac{4 a_0^2 r}{D_T } } \approx \frac{D_A}{D_T} +\frac{a_0^2 r}{D_T}-\frac{2a_0^2 r}{D_A+a_0^2 r}$, $\alpha_1 \approx \frac{1}{a_0^2}\frac{a_0^2 r}{D_A+a_0^2 r}\approx \frac{1}{a_0^2}\frac{1}{\left[1+\frac{\mu}{r}\right]}$ and $\alpha_2 \approx \frac{1}{a_0^2}\frac{D_A}{D_T}\left[1 +\frac{ r}{\mu}\right].$  So  the steady-state distribution (\ref{P_r_ss}) becomes
\begin{align}
P_{r,ss}(x)& \approx \frac{1}{2a_0} \frac{1+\frac{\mu}{r}}{\left(2+\frac{r}{\mu}+\frac{\mu}{r}\right)\sqrt{1+\frac{\mu}{r}}} e^{-\sqrt{\frac{1}{\left[1+\frac{\mu}{r}\right]}}\frac{|x|}{a_0}}\nonumber\\
&+\frac{1}{2  a_0}\frac{1+\frac{r}{\mu}}{\left(2+\frac{r}{\mu}+\frac{\mu}{r}\right) }\sqrt{\frac{D_A}{D_T}\left(1 +\frac{ r}{\mu}\right)} e^{-\sqrt{\frac{D_A}{D_T}\left(1 +\frac{ r}{\mu}\right)}\frac{|x|}{a_0}}.\label{Pr_largemu}
\end{align}
 Here we can consider a limit, $r\gg \mu,$ and $D_T/D_A$ is finite. For such case,
\begin{align}
  P_{r,ss}(x)\approx \frac{1}{2a_0} \frac{\mu}{r} e^{-\frac{|x|}{a_0}}+ \frac{1}{2}\left(1-\frac{\mu}{r}\right)\sqrt{ \frac{r}{D_T}}\,e^{-\sqrt{ \frac{r}{D_T}} |x|}.\label{graph1b}
\end{align}
 In the above equation,  the first term on the RHS can be ignored, and so the distribution is given by $P_{r,ss}(x)\approx  \frac{1}{2}\sqrt{ \frac{r}{D_T}}\,e^{-\sqrt{ \frac{r}{D_T}} |x|}.$  For large values of $r,$ the distribution is mostly peaked around  $x_0=0,$ in accordance with Fig.  \ref{plot_pdfshotr}(b), as the particle is  brought back frequently to the origin and thus it  is mostly found there. For the case where $r< \mu$ and $D_A\gg D_T,$ 
the first term on the RHS in Eq. (\ref{Pr_largemu}) mostly contributes to the tail behavior, and around $x=0$ the second term gives a delta function, $viz.$ 
$ \delta(x)/\left(1 +\frac{\mu}{r}\right).$

\subsection{Different limits of $\left(D_A/D_T\right)$ \label{diffepsilon}}
\begin{figure}
\centering
\begin{subfigure}{.5\textwidth}
  \centering
  \includegraphics[width=1.0\linewidth]{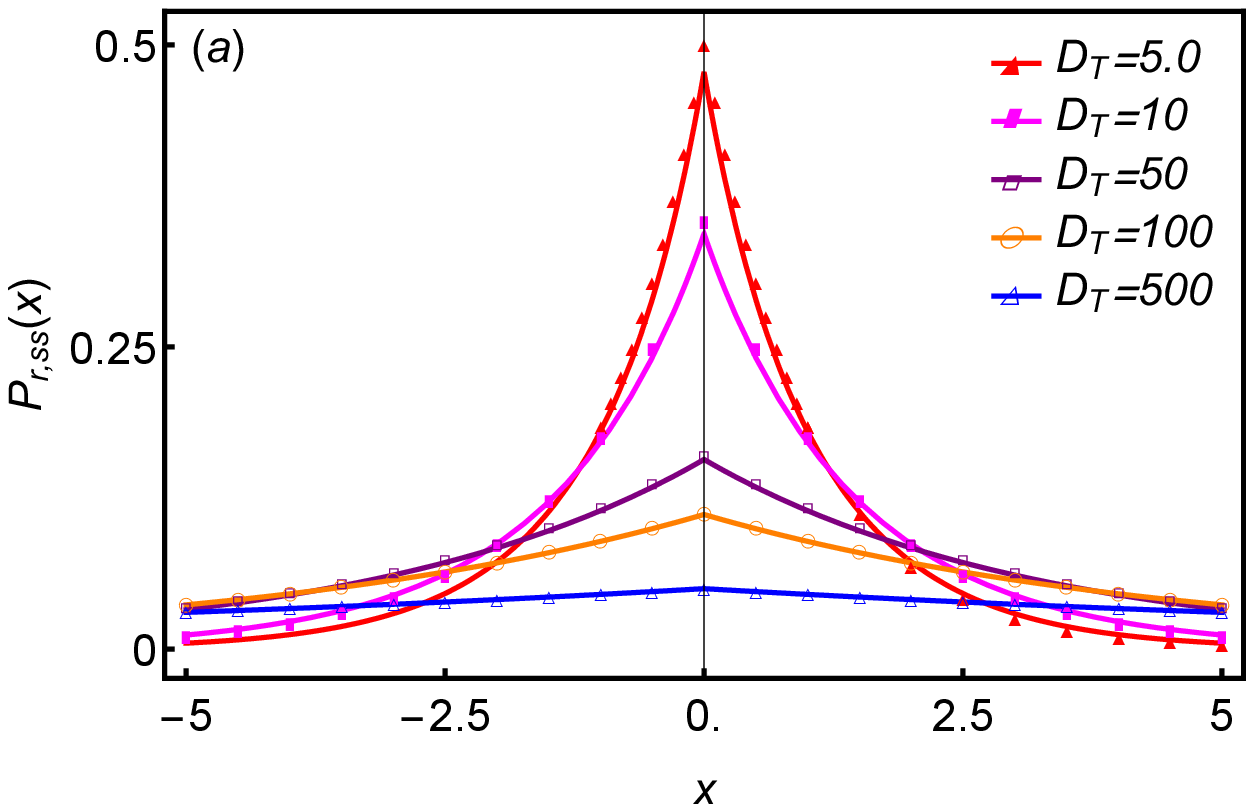}
\end{subfigure}%
\begin{subfigure}{.5\textwidth}
  \centering
  \includegraphics[width=1.0\linewidth]{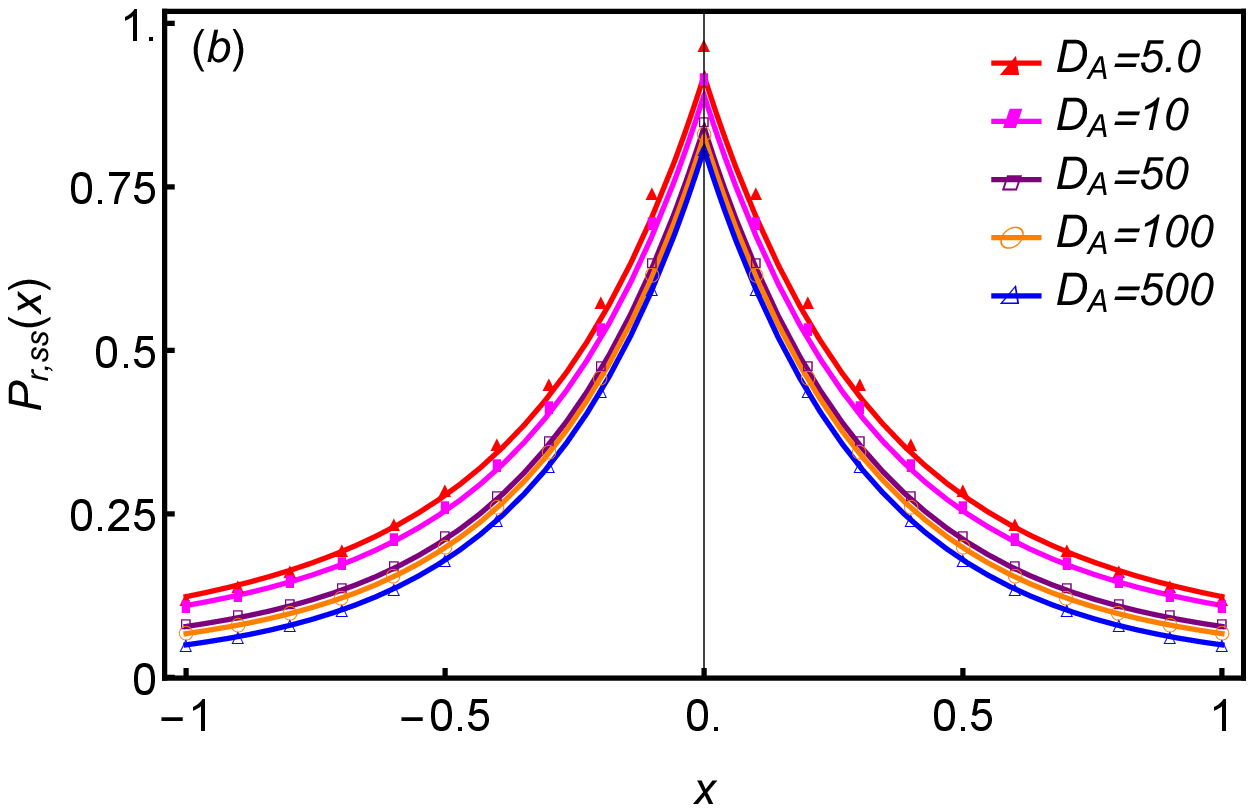}
\end{subfigure}
\caption{Plot of the probability distribution functions [Eq. (\ref{P_r_ss})] as a function of displacements for different values of $D_A/D_T.$ In panel (a) $D_T>D_A,$ where we have taken $D_A=1.0.$  In panel (b) $D_A>D_T,$ where $D_T=1.0.$ Other parameters are $r=\mu=5.0.$ Approximate results of the PDF given in Eqs. (\ref{P_thermal}) and (\ref{p_active}) represented by the lines with point symbols are in good agreement with the exact one (solid lines) as shown in panels (a) and (b), respectively. }
\label{PlotfordifferentD}
\end{figure}

We first consider the limit $\frac{D_A}{D_T} \ll 1,$ where $\sqrt{\left[\frac{D_A}{D_T} +1+\frac{a_0^2 r}{D_T}\right]^2- \frac{4 a_0^2 r}{D_T } } \approx \Big|1-\frac{a_0^2 r}{D_T}\Big|.$ For $\frac{a_0^2 r}{D_T}>1$,
$\alpha_1(r)\approx 1/a_0^2,$ and $\alpha_2(r) \approx \frac{r}{D_T}.$ In the other limit, $\frac{a_0^2 r}{D_T}<1$, $\alpha_1(r)$ and $\alpha_2(r)$ just exchange their values.  Therefore, for both cases,  Eq. (\ref{P_r_ss}) approximates to 
\begin{align}
P_{r,ss}(x)  \approx \frac{1}{2}\sqrt{ \frac{r}{D_T}}\,e^{-\sqrt{ \frac{r}{D_T}} |x|}.\label{P_thermal}
\end{align}
Here, the thermal contribution prevails over the active one, and as expected,  we recover the well-known result for the normal diffusion with resetting. This is consistent with  Fig.
 \ref{PlotfordifferentD} (a).
 
Now we take the limit $\frac{D_A}{D_T} \gg 1.$  So one can do the following approximations:
$\sqrt{\left[\frac{D_A}{D_T} +1+\frac{a_0^2 r}{D_T}\right]^2- \frac{4 a_0^2 r}{D_T } } \approx \frac{D_A}{D_T} +\frac{a_0^2 r}{D_T}-\frac{2a_0^2 r}{D_A+a_0^2 r}$,
 $\alpha_2(r)\approx \frac{1}{a_0^2}\left[\frac{D_A}{D_T} +\frac{a_0^2 r}{D_T}\right],\,\alpha_1(r) \approx \frac{r}{D_A+a_0^2 r}=\frac{1}{a_0^2}\frac{1}{1+\frac{\mu}{r}}.$  Therefore, using Eq. (\ref{P_r_ss}) one  has  
\begin{align}
P_{r,ss}(x) \approx  \frac12\left(\frac{D_A}{D_A+a_0^2r}\right)\sqrt{\frac{r}{D_A+a_0^2r}} e^{-\sqrt{\frac{r}{D_A+a_0^2r}}|x|}+\frac{1}{2a_0}\left(\frac{a_0^2r}{D_A+a_0^2 r}\right) \sqrt{\frac{D_A}{D_T} +\frac{a_0^2 r}{D_T}}\,e^{-\sqrt{\frac{D_A}{D_T} +\frac{a_0^2 r}{D_T}} \frac{|x|}{a_0}}.\label{p_active}
\end{align}
 The first term on the RHS dictates large$-x$ behavior whereas the second one only contributes to the distribution at $x=0.$ Such features can be easily seen  in Fig.
 \ref{PlotfordifferentD} (b).  For $\frac{a_0^2 r}{D_A}\ll 1$ or $\frac{r}{\mu}\ll 1,$ the  PDF  can be approximated as $P_{r,ss}(x) \approx  \frac12\sqrt{\frac{r}{D_A}} e^{-\sqrt{\frac{r}{D_A}}|x|},$ which is of similar form as Eq. (\ref{P_thermal}) with the only difference being that the thermal diffusivity is now replaced  by the active one.   At $\frac{r}{\mu}\gg 1,$ the second term on the RHS of Eq. (\ref{p_active}) is dominant over the first one, and thereby the PDF can be given as $P_{r,ss}(x)\approx  \frac{1}{2}\sqrt{ \frac{r}{D_T}}\,e^{-\sqrt{ \frac{r}{D_T}} |x|} \approx \delta(x).$  Now we consider a situation where only the Poisson noise is present in the system, Therefore, taking  $D_T\rightarrow 0$ in  Eq. (\ref{p_active}) the second term   turns to a delta function, and so Eq. (\ref{p_active}) can be recast as
\begin{align}
P_{r,ss}(x) \approx  \frac{1}{2a_0}\left(\frac{1}{1+\frac{r}{\mu}}\right)\sqrt{\frac{1}{1+\frac{\mu}{r}}} e^{-\sqrt{\frac{1}{1+\frac{\mu}{r}}}\frac{|x|}{a_0}}+\left(\frac{1}{1+\frac{\mu}{r}}\right)\delta(x).
\end{align}

 \section{Relaxation to nonequilibrium steady state \label{relaxation}}
 Here we analyze how the distribution relaxes to the stationary one at long times. For this purpose, we approximate the distribution after doing the inverse Fourier transform of  Eq. (\ref{renewal_Fourier}) as discussed in Appendix. \ref{appen3}. Two limiting cases for the distribution given in  Eq. (\ref{approx_P}) are discussed  in the following sections.
 \subsection{$1>r \gg \mu $ and  $D_T\gg D_A$  }
\begin{figure}
\centering
\begin{subfigure}{.5\textwidth}
\centering
 \includegraphics[width=1.0\linewidth]{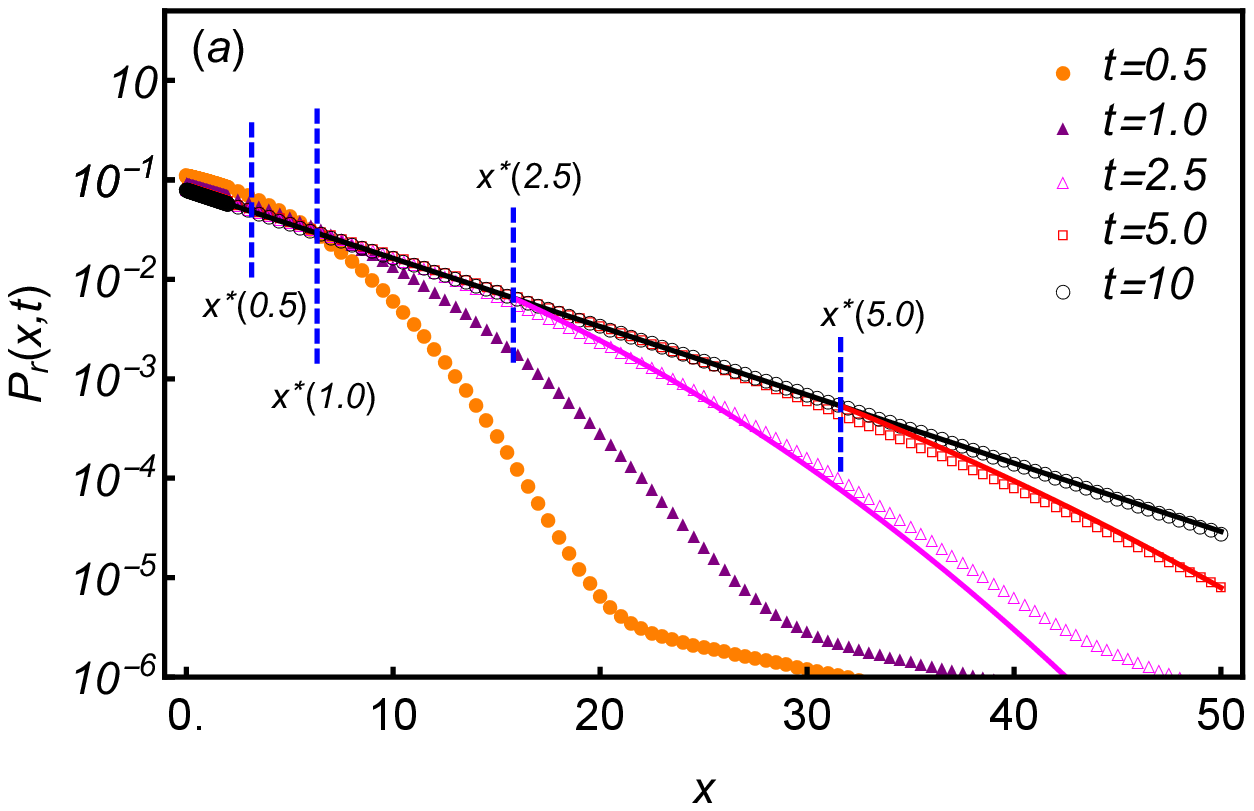}
\end{subfigure}%
\begin{subfigure}{0.5\textwidth}
 \centering
 \includegraphics[width=1.0\linewidth]{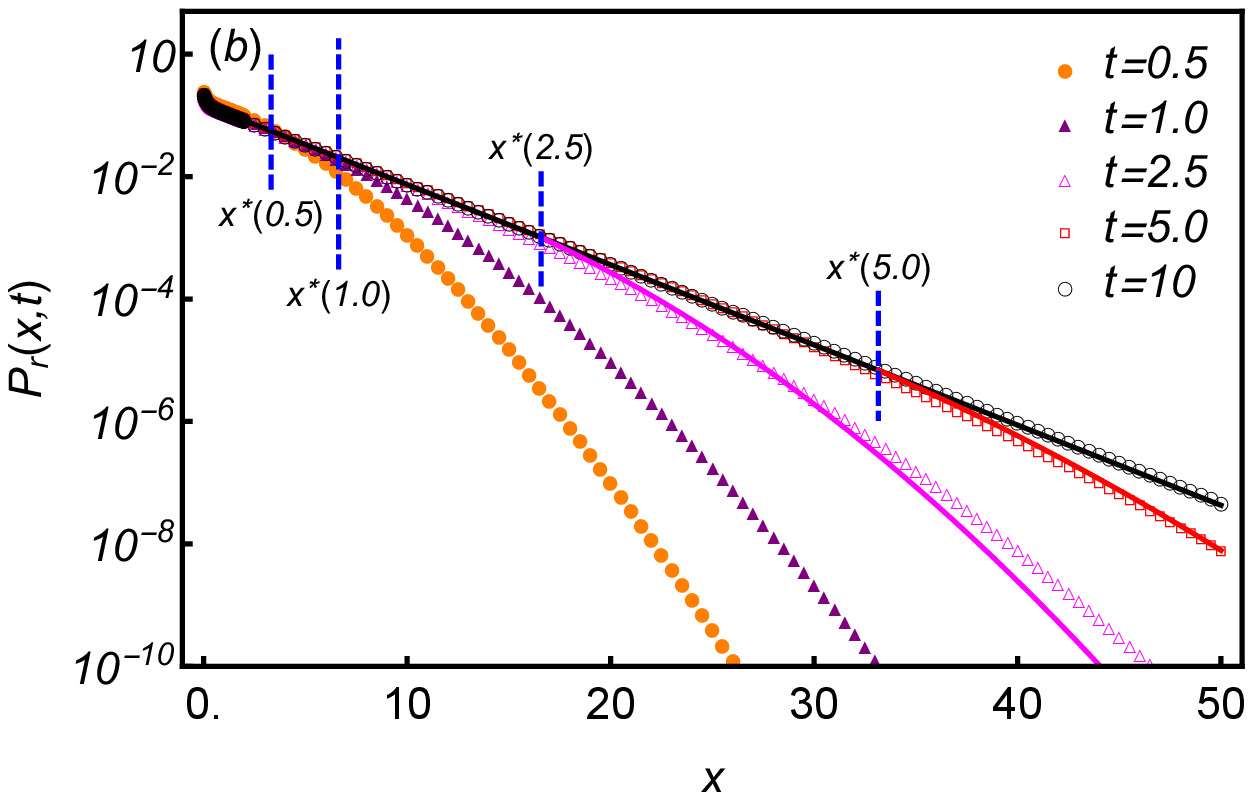}
\end{subfigure}
\caption{Comparative plots of analytical results of  the complete  probability distribution function (PDF) with the numerical one in logarithmic scale as a function of displacements at different times.   The dotted curves are plotted using the data obtained from the numerical inverse Fourier transform  of Eq. (\ref{renewal_Fourier1}). The position $x^*(t)$ which distinguishes two regimes is marked by blue dashed vertical lines for different times. In panels (a) and (b) the plots are for two limiting cases: (a) $\mu \ll 1 $, $D_T\gg D_A$ and (b) $\mu \gg r$, $D_A/D_T$ is finite. The  values of parameters used for computations are given by  (a) $D_T=20,\,D_A=0.1,\,\mu=0.001,\,r=0.5$  and (b) $D_T=1.0,\,D_A=10,\,\mu=40,\,r=1.0.$  The solid lines corresponds to analytical results given in Eqs. (\ref{approx_P_smalle1}) and (\ref{approx_P_largee1}) for two respective  cases. The black solid (upper) line is for $x(t)<x^*(t)$ whereas other solid curves (magenta, red) represent the PDFs for large values of $t$ (here we have taken $t=2.5,\,5.0,\,10$) in the region $x(t)>x^*(t).$  }
\label{Plotfordifferentr}
\end{figure}

  The integration over $\phi$  in Eq. (\ref{approx_P}) can be done using the Laplace method, according to which,  most non-zero contributions should arise near $\phi=0$ for the large value of $t.$ In this limit, $I_1(2\mu t' \phi) \approx \mu t'\phi \rightarrow 0$ at  $\phi = 0.$ For $D_T\gg D_A$ and $1 \gg \mu, $  the second term on the RHS is negligible  compared to the first one. Therefore, $ P_r(x,t)$ can be asymptotically given as
$$  P_r(x,t) \approx  r\int_{0}^{t}\, dt'\,e^{-r t'-\mu\,t'} \frac{e^{-\frac{x^2}{4 D_T t'}}}{\sqrt{4 \pi D_T t'}}.$$ Taking $t'=\omega t,$ the above equation can be recast as 
\begin{align}
P_r(x,t) \approx  \frac{ rt^{1/2}}{\sqrt{4 \pi D_T }}\int_{0}^{1}\, \frac{d\omega}{\sqrt{\omega}}\,e^{-t \Omega_1(\vrv,\omega)},\label{approx_P_smalle}
\end{align}
where the large deviation function (LDF) is,   $\Omega_1(\vrv,\omega)=r \omega+\mu \omega+\frac{\vrv^2}{4 D_T\omega} \approx r \omega+ \frac{\vrv^2}{4 D_T\omega}$ for  $r\gg \mu,$ and $\vrv=x/t.$ For large value of $t,$ the integration over $\omega$ can be performed using the saddle-point method. Most contributions to the integration comes where $\Omega_1'(\vrv,\omega^*)=0,$ which  means $\omega^*= \frac{|\vrv|}{2\sqrt{D_T r}}.$   One can distinguish two distinct regimes across $\omega^*=1$ as illustrated in Ref \cite{evans2020stochastic}. The minimum  of  $\Omega_1(\vrv,\omega)$ occurs (i) at  $\omega^*$ in the region $\omega\in [0,1),$ and (ii) at $\omega^*=1$ for $\omega^*>1.$  So the distribution can be described by the following LDFs at large $t$:
\begin{align}
\Omega_1(\vrv,\omega)=
\begin{cases}
\sqrt{\frac{r}{D_T}}|\vrv|,&  \text{for}\, |\vrv|<\vrv^*\\
r+\frac{\vrv^2}{4D_T},&    \text{for}\, |\vrv|>\vrv^*
\end{cases},
\end{align}
where $\vrv^*=2\sqrt{D_T\,r},$ or   equivalently, one can define an associated length  scale $x^*$ as $x^*(t)=2\sqrt{D_T r}\, t.$ So, in the spatial region $|x|<x^*,$  the density relaxes to a steady state, but it is still in the transient regime for $|x|>x^*.$  This result  is  equivalent to  one in the normal case, as has been found in Sec. \ref{diffepsilon}. Doing the integration in Eq. (\ref{approx_P_smalle}) and  keeping all the terms carefully, we find  the pre-exponential factor to be $\frac{r \omega^*}{|\vrv|}.$ Therefore, the complete PDF in the entire space can be expressed as 
\begin{align}
P_r(x,t)=\Theta(x^*-|x|)\frac{1}{2}\sqrt{\frac{r}{D_T}}\,e^{-\sqrt{\frac{r}{D_T}} |x|}+\Theta(|x|-x^*)\frac{r t}{|x|}\,e^{-r t -\frac{x^2}{4 D_T t}},\label{approx_P_smalle1}
\end{align}
 where $\Theta(z)$ is the   Heaviside step function. The analytical result (\ref{approx_P_smalle1}) agrees well with the numerical one shown in Fig. (\ref{Plotfordifferentr}) (a).

 \subsection{$\mu \gg r \geq 1$ and $D_A/D_T$ is finite }
For large values of $t$ and $\mu,$ the first term on the RHS of Eq. (\ref{approx_P}) is negligible compared to the second one.  Applying the asymptotic form of $I_1(z)$ as mentioned in Appendix \ref{appen3}, the second term can be approximated as 
\begin{align}
  P_r(x,t)\approx 2 r\int_{0}^{t}\,(\mu t') dt'\,\int_{0}^{\infty}d\phi\,e^{-r t'}\,\frac{e^{-\mu t'( \phi-1)^2}}{\sqrt{4\pi\mu t' \phi}}\frac{e^{-\frac{x^2}{4 t'(D_T+D_A \phi^2)}}}{\sqrt{4 \pi t'(D_T+D_A \phi^2)}}.\label{Pr_approx}
   \end{align}
 Now using the saddle-point approximation, the integration over $\phi$ can be done as  most contributions come from the saddle-point at $\phi=1.$ 
  Equation (\ref{Pr_approx}) approximates to
 \begin{align}
P_r(x,t)\approx  \frac{ rt^{1/2}}{\sqrt{4 \pi (D_A+D_T)}}\int_{0}^{t}\,dt'\,\frac{e^{-\frac{x^2}{4(D_A+D_T) t'}}}{\sqrt{ t'}}.
 \end{align}
 For large $t,$  taking  $t'=\omega t,$ one obtains
  \begin{align}
  P_r(x,t)& \approx \frac{ rt^{1/2}}{\sqrt{4 \pi (D_A+D_T)}}\int_{0}^{1}\frac{d\omega}{\sqrt{\omega}}\,e^{-t \Omega_2(\vrv,\omega)},     
  \end{align}
 where the LDF is given by $\Omega_2(\vrv,\omega)=r\omega+\frac{\vrv^2}{4 (D_A+D_T)\omega},$  which implies,  $\omega^*=\frac{|\vrv|}{2\sqrt{(D_A+D_T) r}}.$ Like the previous case, we can identify two regimes as follows: 
 \begin{align}
\Omega_2(\vrv,\omega)=
\begin{cases}
   \sqrt{\frac{r}{D_A+D_T}}|\vrv|,&  \text{for}\, |\vrv|<\vrv^*\\
    r+\frac{\vrv^2}{4(D_A+D_T)},&    \text{for}\, |\vrv|>\vrv^*
\end{cases},
\end{align} 
 where $\vrv^*=2\sqrt{(D_A+D_T) r},$ or $x^*=2\sqrt{(D_A+D_T) r} \,t.$ So putting all terms together, we can write the PDF as 
\begin{align}
P_r(x,t)=\Theta(x^*-|x|)\frac{1}{2}\sqrt{ \frac{r}{D_A+D_T}}\,e^{-\sqrt{ \frac{r}{D_A+D_T}} |x|}+\Theta(|x|-x^*)\frac{r t}{|x|}\,e^{-r t -\frac{x^2}{4 (D_A+D_T) t}}.\label{approx_P_largee1}
\end{align}
In Fig. (\ref{Plotfordifferentr}) (b), Eq. (\ref{approx_P_largee1}) is compared with the PDF computed numerically  using  Eq. (\ref{renewal_Fourier1}), and they are in a good agreement for $t>1.$ The position $x^*$  dictates the partition between two domains, $|x|<x^*$ and $|x|>x^*,$  and it moves linearly with time $t.$ As  time progresses, the first domain merges to the second one, and thus the steady state  is established in the entire space.  For $D_A\gg D_T,$ the distribution at the steady state can be computed using Eq. (\ref{approx_P_largee1}), as $P_{r,ss}(x) \approx  \frac12\sqrt{\frac{r}{D_A}} e^{-\sqrt{\frac{r}{D_A}}|x|},$ and the speed at which the first domain grows mostly depends on $D_A,$ or more specifically,  the system reaches quickly to the steady state as  the strength of nonthermal fluctuations increases. 

\section{First-arrival problem  \label{First-passage-properties}}

Here we address the first-arrival problem for a  particle diffusing in the presence of a target at position $x=x_s.$  The particle was initially at  position $x=x_0(\neq x_S),$ and it is under the Poissonian resetting mechanism which resets the particle to position $x_0$ at random times with a fixed rate $r.$ The probability of finding  the particle diminishes when  it encounters the target, which can be thought of as a trap and  is modeled here by  a delta-function sink of strength $\kappa$ \cite{szabo1984, bagchi1987jcp, bagchi1990jpc,sebastian1992, chakrabartijcp2006}. The particle disappears on arriving at the sink  if   the value of $\kappa$ goes to infinity. The time it takes to reach the sink is a random variable, and for the  reset-free case in a   semi-infinite region, the average time is infinite.  But in the presence of resetting, the time is usually finite and takes a minimum value at some resetting rate $r^*.$ Clearly this  leads to a more efficient search strategy. For  finite values of  $\kappa,$  the particle is not completely absorbed on its first  arrival  at $x=x_s$ (partial absorption), and so it visits the sink multiple times throughout its journey  before it gets annihilated at the sink.  As mentioned in Sec. \ref{intro}, there are several cellular processes where reactants cannot recognize the target site at  all times. For instance, the transcription is initiated when the protein called the  transcription factor (TF) binds to a promoter site on the DNA backbone, but depending on the folding state of chromatin, the site becomes ``visible" or ``hidden" to the protein. Only in the visible state, the binding happens \cite{munsky2012using,mercado2019first}. Such reactions can be  conceived as a form of  the partial absorption in the presence of a delta sink with finite $\kappa.$

 We are dealing here with the process which is Markovian in nature, $i.e.,$ the absence of any memory into the dynamics.  For such processes, one can find the first-arrival time density  using the Green's-function method as detailed in \cite{szabo1984, sebastian1992}. Without any resetting, the first-arrival time density (FATD) is denoted as $f_{T_0}(t),$ and its Laplace transform is given by  $\Tilde{f}_{T_0}(s)=\int_{0}^{\infty} dt\,e^{-s t}\,f_{T_0}(t).$  So in the presence of a sink of strength $\kappa,$ the FATD in the Laplace domain  can be expressed as \cite{sebastian1992,PhysRevE.95.012154}
 \begin{align}
\Tilde{f}_{T_0}(s)=\frac{\kappa\,\Tilde{P}_0(x_s,s|x_0)}{1+\kappa\Tilde{P}_0(x_s,s|x_s)},\label{FPTk0}
 \end{align} 
 where $P_0(x,t|x_0)$  is the free propagator,   given by 
$
P_0(x,t|x_0)=\int \frac{dp}{2\pi} \,e^{-i p (x-x_0)}e^{-\left(D_T p^2\,t + D_A\,t \frac{p^2}{1+ p^2} \right)}
$ [see Eq. (\ref{FP0_fourier})],
and its Laplace transform is (see Eq. (\ref{P_r_ss}))
\begin{align}
  \Tilde{P}_0(x,s|x_0)&=\frac{P_{r,ss}(x)\Big|_{r=s}}{r}=\int_{0}^{\infty}dt\,e^{-s t}P_0(x,t|x_0)\nonumber\\
  &=\left(\frac{\alpha_3(s)-\alpha_1(s)}{\alpha_2(s)-\alpha_1(s)}\right) \frac{e^{-\sqrt{\alpha_1(s)} |x-x_0|}}{2D_T\sqrt{\alpha_1(s)}}+\left(\frac{\alpha_2(s)-\alpha_3(s)}{\alpha_2(s)-\alpha_1(s)}\right) \frac{e^{-\sqrt{\alpha_2(s)} |x-x_0|}}{2D_T\sqrt{\alpha_2(s)}}.\label{G0}
\end{align}

Let us now invoke  the resetting mechanism to  the given problem as described earlier.  With resetting, the FATD in the Laplace domain can be  expressed in terms of  $\Tilde{f}_{T_0}(s)$ given in Eq. (\ref{FPTk0}), as \cite{reuveni2016optimal}
 \begin{align}
\Tilde{f}_{T_r}(s)=\frac{\Tilde{f}_{T_0}(s+r)}{\frac{s}{s+r}+\frac{r}{s+r}\Tilde{f}_{T_0}(s+r)}.\label{FPTr}
 \end{align}

An important quantity for a search process is  the mean first-arrival time (MFAT) which is basically the first moment of  FATD. So it can be easily calculated using the following expression: 
 \begin{align}
 \langle \mathbb{T}_r \rangle =-\frac{\partial\Tilde{f}_{T_r}(s)}{\partial s}|_{s=0}=\frac{1}{r}\frac{1}{\Tilde{f}_{T_0}(r)}-\frac{1}{r}=\frac{1}{r}\frac{1+\kappa\Tilde{P}_0(x_s,r|x_s)}{\kappa\Tilde{P}_0(x_s,r|x_0)}-\frac{1}{r}.\label{MFPTr}
 \end{align}
 
 \begin{figure}
\centering
\begin{subfigure}{.5\textwidth}
  \centering
  \includegraphics[width=1.0\linewidth]{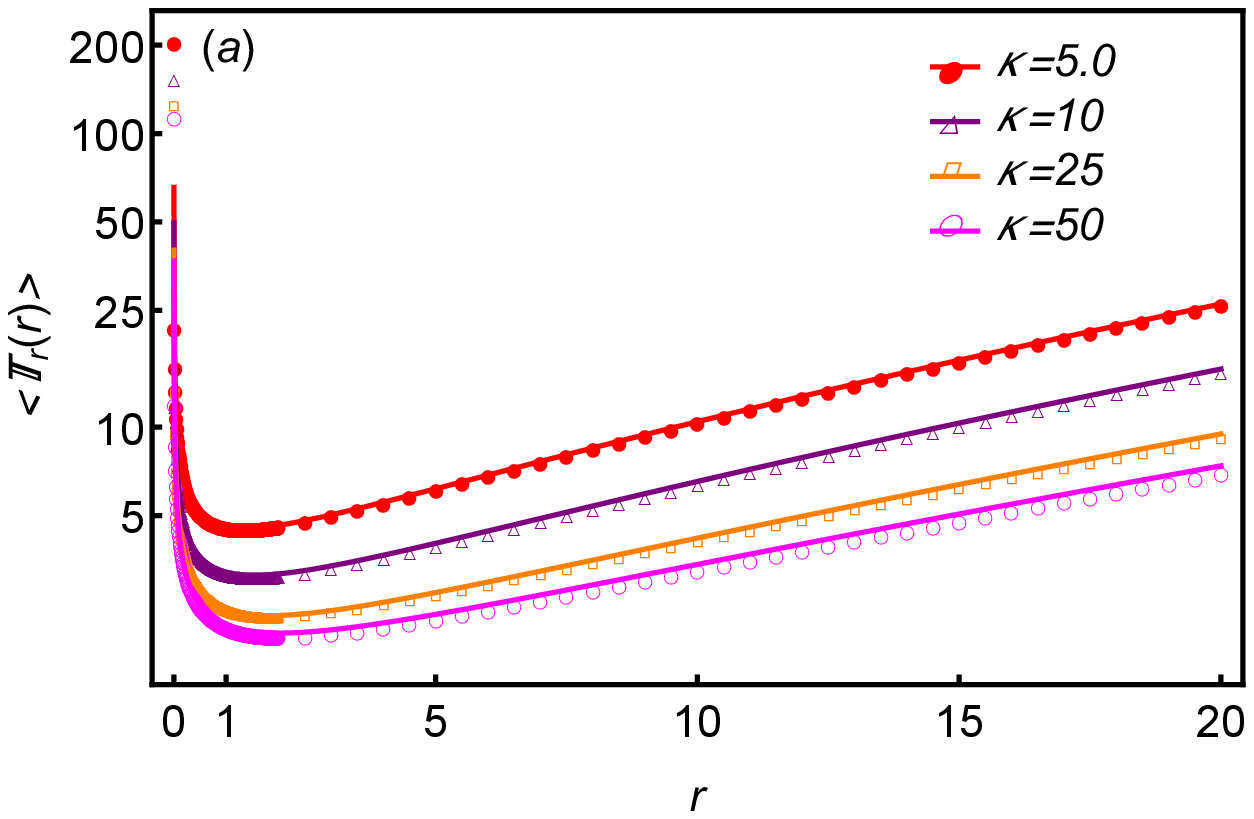}
\end{subfigure}%
\begin{subfigure}{.5\textwidth}
  \centering
  \includegraphics[width=1.0\linewidth]{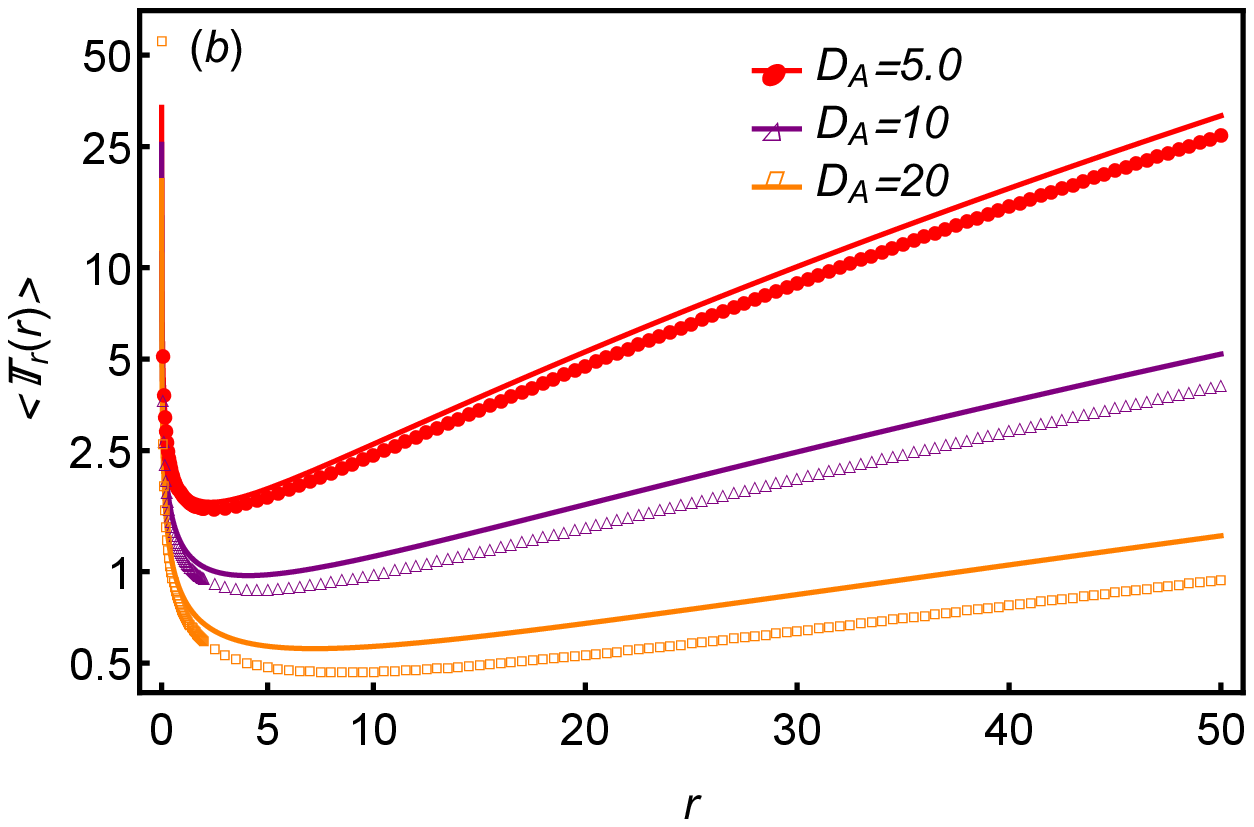}
\end{subfigure}
\caption{Logarithmic plot of first-passage times versus resetting rate $r:$ (a) for different values of sink strength $\kappa$ taking $D_A=5.0,\,D_T=1.0$ and (b)  for different $D_A$ values keeping the thermal diffusivity constant at $D_T=1.0$ in the case of complete absorption. The solid curves are drawn  by solving  Eq. (\ref{MFPTr}) numerically taking $\mu=5000$ and assuming that the distance between the sink and the source is fixed at $|x_s-x_0|=2.50.$ The dotted curves correspond to Eq. (\ref{MFAT_smallr}) which is a quite approximate result for the limit $\frac{D_A}{D_T}\frac{r}{\mu}\ll 1.$ In panel (a) the solid curves agree well with the analytical results for the given range of Poisson rate $r,$ whereas the dotted lines are good fits only  for small values of  $r$ if $D_A$ is large, as shown in panel (b).  }
\label{plot_MFAT_vs_k}
\end{figure}
Employing  the analysis done in Sec. \ref{r limit} (for the limiting case $\frac{a_0^2r}{D_T}\ll 1$),  one can obtain, for small values of $r,$ $$\Tilde{P}_0(x_s,r|x_0) \approx \frac{1}{2 r} \sqrt{\frac{r}{D_T+D_A}}e^{-\sqrt{\frac{r}{D_T+D_A}}|x_s-x_0|}.$$  Using the above, the MFAT can be approximated as 
\begin{align}
 \langle \mathbb{T}_r \rangle \approx \frac{1}{r}\frac{2r\sqrt{\frac{D_T+D_A}{r}}+\kappa}{\kappa\,e^{-\sqrt{\frac{r}{D_T+D_A}}|x_s-x_0|}}-\frac{1}{r} \approx  \frac{1}{r}\left(e^{\sqrt{\frac{r}{D_T+D_A}}|x_s-x_0|}-1\right)+\frac{2}{\kappa}\sqrt{\frac{D_T+D_A}{r}}\,e^{\sqrt{\frac{r}{D_T+D_A}}|x_s-x_0|}.\label{MFAT_smallr}
\end{align}
At $r\rightarrow 0$, $\langle \mathbb{T}_r \rangle \approx \sqrt{\frac{1}{r(D_A+D_T)}}|x_s-x_0|+\frac{2}{\kappa}\sqrt{\frac{D_T+D_A}{r}}+\frac{2}{\kappa}|x_s-x_0|.$ In the case of  complete absorption ($\kappa \rightarrow \infty$), the MFAT can be written as $\langle \mathbb{T}_r \rangle \approx \sqrt{\frac{1}{r(D_A+D_T)}}|x_s-x_0|.$ As expected, for no resetting case,  $\langle \mathbb{T}_r \rangle$  diverges as $r^{-1/2}$ as $r \rightarrow 0.$
For the case where only the thermal noise acts on the particle, $i.e.$, $D_A\rightarrow 0,$ the MFAT given in Eq. (\ref{MFAT_smallr})  transforms to
\begin{align}
 \langle \mathbb{T}_r \rangle \approx \frac{1}{r}\left(e^{\sqrt{\frac{r}{D_T}}|x_s-x_0|}-1\right)+\frac{2}{\kappa}\sqrt{\frac{D_T}{r}}e^{\sqrt{\frac{r}{D_T}}|x_s-x_0|},\label{MFAT_smallrdt}
\end{align}
which is equivalent to the one given in Ref. \cite{whitehouse2013effect}.   For $D_A\gg D_T$ and small values of $r,$ 
\begin{align}
 \langle \mathbb{T}_r \rangle \approx \frac{1}{r}\left(e^{\sqrt{\frac{r}{D_A}}|x_s-x_0|}-1\right)+\frac{2}{\kappa}\sqrt{\frac{D_A}{r}}e^{\sqrt{\frac{r}{D_A}}|x_s-x_0|}.\label{MFAT_smallrda}
\end{align}
For extremely large values of $r$ and finite values of $D_A$ and $D_T,$  we can have
$\Tilde{P}_0(x_s,r|x_0) \approx\frac{1}{2r}\sqrt{ \frac{r}{D_T}}\,e^{-\sqrt{ \frac{r}{D_T}} |x_S-x_0|},$ and therefore, the MFAT can be described by Eq. (\ref{MFAT_smallrdt}).

\begin{figure}
\centering
\begin{subfigure}{0.4\textwidth}
  \centering
  \includegraphics[width=1.0\linewidth]{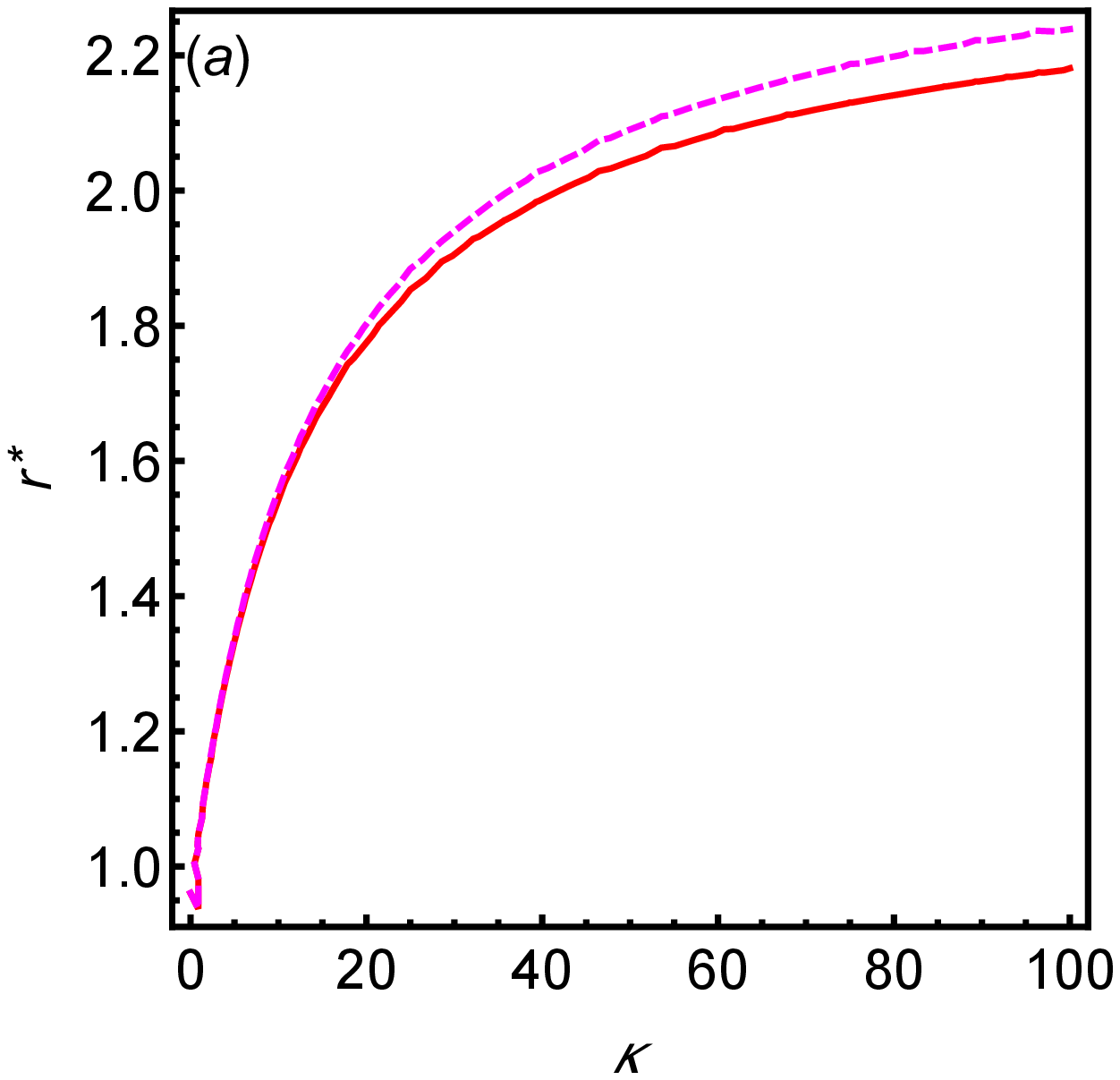}
\end{subfigure}%
\begin{subfigure}{0.4\textwidth}
  \centering
 \includegraphics[width=1.0\linewidth]{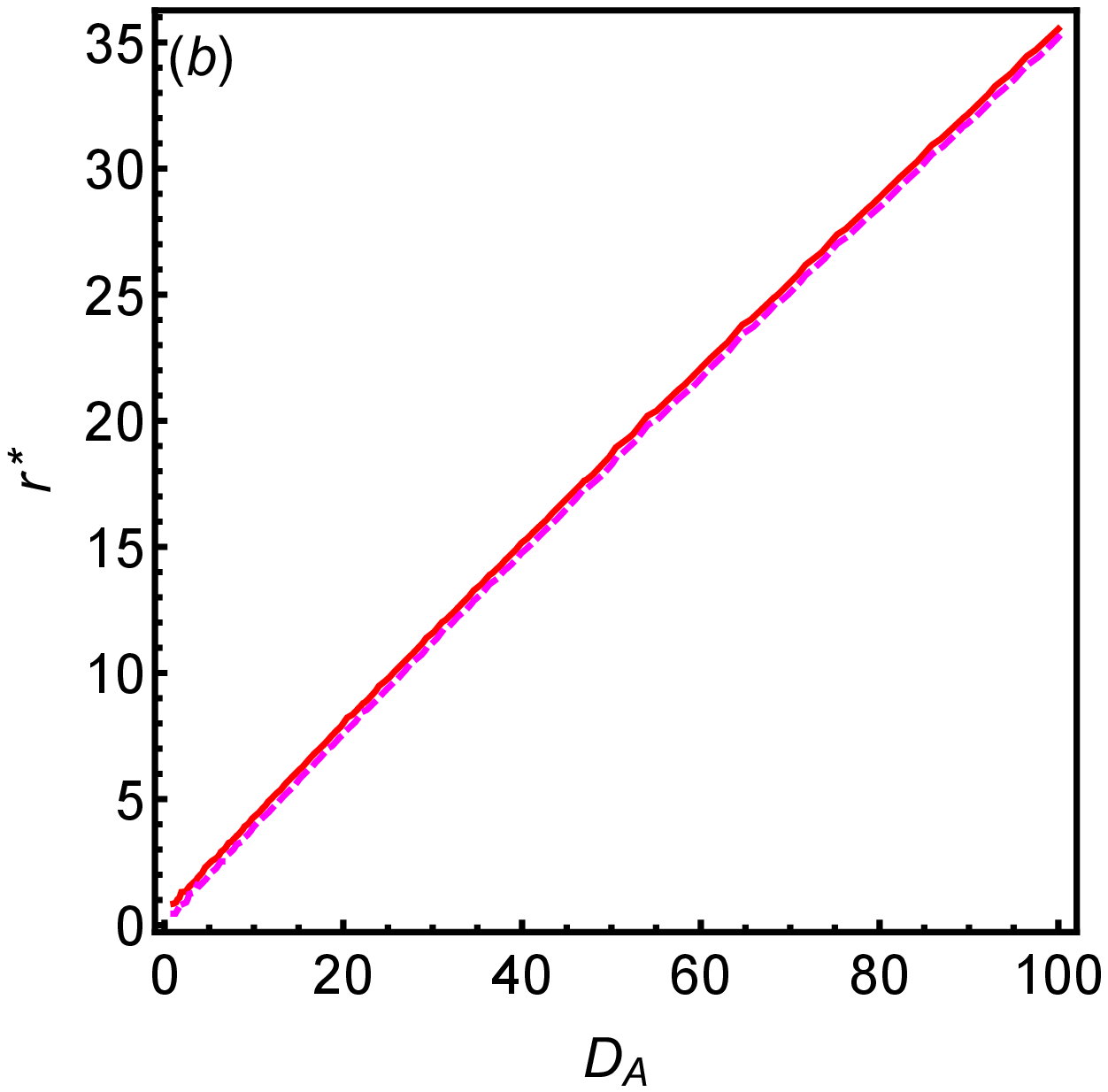}
\end{subfigure}
\caption{(a)  Plot of optimal rate $r^*$ as a function of the sink strength $\kappa.$ The other parameters are  $D_A=5.0,\,D_T=1.0.$  (b) The optimal rate $r^*$ for complete absorption is plotted against    $D_A$ considering $D_T$ constant at $D_T=1.0.$ The curves are obtained  by solving the equation: $\frac{d}{dr}\langle \mathbb{T}_r \rangle|_{r=r^*}=0.$  Solid lines correspond to the exact results computed with the aid of Eqs. (\ref{MFPTr}) and (\ref{G0}), and approximate result given in Eq. (\ref{MFAT_smallr}) is used to draw the dashed lines. For both panels  (a) and (b), the other parameters taken for numerical calculation are given here:    $|x_S-x_0|=2.50,\,a_0=0.02236.$ }
\label{rate_vs_kappa-epsilon}
\end{figure}

 Figure \ref{plot_MFAT_vs_k} (a) demonstrates results for the MFAT   as a function of resetting rate $r$ for different values of $\kappa.$ In the limits, $r\rightarrow 0$ and $r\rightarrow \infty$, $\langle \mathbb{T}_r \rangle$ blows up to infinity as can be understood from  Figure \ref{plot_MFAT_vs_k}. This can be explained as follows: for $r\rightarrow 0,$ some trajectories can never cross the sink due to almost no resetting, and this leads to infinite MFAT.  On the other hand, if $r\rightarrow \infty,$ the particle is reset so frequently that it always remains near the initial position and so  it never reach the sink. In between these extreme limits, $\langle \mathbb{T}_r \rangle$ varies nonmonotonically, namely,  the MFAT decreases with $r$  until it reaches to a  minimum. So there exists an optimal resetting rate $r^*$ for which one has,  $\frac{d}{dr}\langle \mathbb{T}_r \rangle|_{r=r^*}=0.$ One needs to solve the previous transcendental equation which, unfortunately, cannot be done analytically. For different values of $\kappa$ and $D_A,$ the optimal rates $r^*$ have been computed numerically and shown in Fig. \ref{rate_vs_kappa-epsilon}.  From Fig. \ref{rate_vs_kappa-epsilon} (a), one can  notice that  $r^*$  increases with  $\kappa,$ but it reaches to a fixed value in the large$-\kappa$ limit which basically corresponds to the complete absorption. For the case of the partial absorption ( $i.e.$, $\kappa$ is small) the particle can easily jump over the sink without getting absorbed, and thus it can avail the entire space. This  suggests that the particle can reach the target from both sides in one dimensional space.  However, for large $\kappa,$ the absorption happens mostly from one side, and thus the process is required to reset more frequently in order to achieve an optimal rate.  Fig.  \ref{plot_MFAT_vs_k} (a) shows the plot of the MFAT as a function of sink strength. Notice that the MFAT is lower for large values of  $\kappa$  as the chances of survival for the particle decrease if the sink has a higher strength.
 
In Fig.  \ref{plot_MFAT_vs_k} (b), it has been shown  that the MFAT, as usual, is a nonmonotonic function of $r$ for any value  of $D_A.$ But for a bath with a fixed thermal diffusivity, the MFAT decreases if the strength (or diffusivity) of athermal noise is large. This is  easy to understand  because these extra fluctuations push the particle away from its initial position and thus help to locate the target quickly. However, for such  acceleration of  the process, one has to bear the cost of its opposing mechanism, $i.e.,$ Poissonian resetting. This means that the density which is transferred  far away from the sink in the opposite direction due to robust diffusive mechanism needs to be brought back to the initial position repeatedly so that  the particle can restart its journey towards the sink. So the optimal rate $r^*$ is an increasing function of $D_A,$ as pictorially depicted in Fig. \ref{rate_vs_kappa-epsilon} (b).

\section{Conclusion\label{conclusion}}
We have investigated the effect of Poissonian resetting on  a diffusing particle in a non-thermal bath. In contrast to the thermal case, the steady-state distribution is described by two exponential functions- one for the central region and another depicting the tail behavior. At transient periods, we have identified two distinct regimes, and have shown how the transient regime gradually relaxes to the stationary one over time,  and the speed at which the relaxation occurs depends on the strengths of the  noises. In the presence of a target, it has been found that the existence of additional noise aids the particle to find the target easily, although in order to get an optimal search rate, the particle is required to reset more frequently. 

Searching  under a resetting protocol  appears to be an effective strategy for a plethora of systems \cite{evans2020stochastic}. In  particular, for biological systems where a reaction  often   starts after a reactant finds a target, the invocation of the resetting mechanism to a reaction becomes very much relevant. Our study essentially focuses on similar aspects where the first arrival of a particle to the sink denotes the beginning of a reaction  and its reversibility is captured via the sink strength. Therefore, our findings bring out a generic perspective of any Markovian  biophysical processes under resetting. However, for a non-Markovian process ($e.g.,$ see Refs. \cite{physRevLett.116.248301,gladrow2017nonequilibrium,chaki2019enhanced}), one requires a different formalism which can be investigated in future.

\section{Acknowledgements\label{acknowledements}}
The authors are thankful to Subhasish Chaki for reading and commenting  on the manuscript. R. C. acknowledges SERB for funding (Project No. MTR/2020/000230 under
 MATRICS scheme) and IRCC-IIT Bombay (Project No. RD/0518-IRCCAW0-001) for funding. K.G.  acknowledges IIT Bombay for support through the institute  postdoctoral fellowship.

\appendix
\section{Derivation of Eq. (\ref{FP0_fourier})\label{Appen1}}
Poisson white noise (PWN) can be realized as sum of pulses $g(t_i)$ with amplitude $a_i$ at different times $t_i$ as:
\[
\xi_A(t)=\sum_{i}a_{i}\, g(t-t_{i})
\]
The number of delta pulses $n$ over a time period $[0,t]$ are drawn from the  Poisson distribution with a  Poisson rate $\mu$, $viz.,$
\begin{equation}
\mathcal{P}(n,t;\mu)=\sum_{n=0}^{n=\infty}\frac{(\mu t)^n}{n!}e^{-\mu t}.
\end{equation}
For such noise, the characteristic functional
can be expressed as \cite{feynman2010quantum}
\begin{equation}
<e^{i\int_{0}^{t}dt_1\, p(t_1)\xi_A(t_1)}>_{\xi_A(t_1)}
=e^{-\mu\int_{0}^{t}dt_1(1-W[\int_{0}^{t_1}dt_2\, p(t_2)g(t_1-t_2)])}\label{characteristic},
\end{equation},
where $W[\psi]=\int_{-\infty}^{+\infty}da\, P(a)e^{ia\psi}.$

The probability distribution of $a$ is $P(a),$ and if we consider
$P(a)$ as the Laplace distribution with scale parameter $a_{0} (>0)$, $i.e.$, $P(a)=\frac{1}{2a_0}e^{-\frac{|a|}{a_0}},$ 
we get,  $W[\psi]=\frac{1}{2a_0}\int_{-\infty}^{+\infty}da\, e^{-\frac{|a|}{a_0}}\,e^{ia\psi}=\frac{1}{1+a_0^2 \psi^2}.$ Using this in Eq. (\ref{characteristic}), one 
obtains
\begin{equation}
<e^{i\int_{0}^{t}dt_1\, p(t_1)\xi_A(t_1)}>_{\xi_A(t_1)}=e^{-\mu\int_{0}^{t}dt_1\frac{a_{0}^{2}(\int_{0}^{t_1}dt_2\, p(t_2)g(t_1-t_2))^{2}}{1+a_{0}^{2}\,(\int_{0}^{t_1}dt_2\, p(t_2)g(t_1-t_2))^{2}}}\label{eq:ch1}.
\end{equation}
Now taking $g(t)=\delta(t),$ one has $\int_{0}^{t_1}dt_2\, p(t_2)g(t_1-t_2)=p(t_1),$ and so Eq. (\ref{eq:ch1}) becomes
\begin{equation}
<e^{i\int_{0}^{t}dt_1\, p(t_1)\xi_A(t_1)}>_{\xi_A(t_1)}=e^{-\mu\int_{0}^{t}dt_1\frac{a_{0}^{2}p(t_1)^{2}}{1+a_{0}^{2}\,p(t_1)^{2}}}\label{eq:ch2}
\end{equation}
 For our model, the dynamical equation reads 
 \begin{align}
     \dot{x}(t)=\eta_T(t)+\xi_A(t),\label{LE}
 \end{align}
 which means that we can write the Fourier transform of the PDF as
 \begin{align}
     \Tilde{P}_0(p,t)=\langle e^{i p x}\rangle=\langle e^{i p \int_{0}^{t}dt_1  \eta_T(t_1)}\rangle_{\eta_T(t_1)}\,\langle e^{i p \int_{0}^{t}dt_1  \xi_A(t_1)}\rangle_{\xi_A(t_1)}
 \end{align}
For thermal noise, $\langle e^{i p \int_{0}^{t}dt_1  \eta_T(t_1)}\rangle_{\eta_T(t_1)}=e^{-D_T p^2\int_{0}^{t}dt_1 }=e^{-D_T p^2 t},$ and for noise $\xi_A(t),$ by virtue of Eq. (\ref{eq:ch2}) one has 
\begin{align}
  \langle e^{i p \int_{0}^{t}dt_1  \xi_A(t_1)}\rangle_{\xi_A(t_1)}=e^{-\mu t \frac{a_0^2 p^2}{1+a_0^2 p^2}}. \label{fp_pwn}
\end{align}
So one can write the Fourier transform as 
\begin{align}
\Tilde{P}_0(p,t)=e^{-D_T p^2 t-\mu t \frac{a_0^2 p^2}{1+a_0^2 p^2}},
\end{align}
which is the same as Eq. (\ref{FP0_fourier}) $(D_A=\mu a_0^2)$. Note that, in Sec. \ref{model} we get the result solving the  Fokker-Planck equation (\ref{FP0}) which is equivalent to the Langevin equation (\ref{LE}).
Taking the derivative of the above with respect to time $t,$ one has 
\begin{align}
\frac{\partial}{\partial t}\tilde{P}_0(p,t)& =-\left[D_T p^2+\mu  \frac{a_0^2 p^2}{1+a_0^2 p^2}\right]\tilde{P}_0(p,t)\nonumber\\
&=-D_T p^2 \tilde{P}_0(p,t)-\mu a_0^2 p^2 \sum_{n=0}^{\infty}(-a_0^2 p^2)^n \tilde{P}_0(p,t)
\end{align}
On doing the inverse Fourier transform, we have 
\begin{align}
\frac{\partial}{\partial t}P_0(x,t)&=D_T \frac{\partial^2  }{\partial x^2}P_0(x,t)+D_A \frac{\partial^2}{\partial x^2}\sum_{n=0}^{\infty}(a_0^2)^n(\frac{\partial^2}{\partial x^2})^n P_0(x,t)\nonumber\\
&=D_T \frac{\partial^2  }{\partial x^2}P_0(x,t)+D_A \frac{\frac{\partial^2}{\partial x^2}}{1-a_0^2 \frac{\partial^2}{\partial x^2}}P_0(x,t),
\end{align}   
 which  is the same as Eq. (\ref{FP0}).

\section{Solving Eq. (\ref{renewal_Fourier})\label{appen2}}

To solve Eq. (\ref{renewal_Fourier}), first Eq. (\ref{renewal_Fourier0})  is  rewritten as
\begin{align}
  &\frac{\partial }{\partial t}\Tilde{P}_r(p,t)+\left[D_T    p^2+D_A \frac{p^2}{1+a_0^2 p^2}+r\right] \Tilde{P}_r(p,t) =r e^{-i p x_0}.\nonumber\\
  &\text{Multiplying  both sides by the integrating factor,}\nonumber\\
 &e^{r t +D_T t  p^2+D_A t \frac{p^2}{1+a_0^2 p^2}}\frac{\partial }{\partial t}\Tilde{P}_r(p,t)+e^{r t +D_T t  p^2 + D_A t \frac{p^2}{1+a_0^2 p^2}}\left[D_T    p^2+D_A \frac{p^2}{1+a_0^2 p^2}+r\right] \Tilde{P}_r(p,t) \nonumber\\
 &=re^{r t  + D_T t  p^2 + D_A t \frac{p^2}{1+a_0^2 p^2}} e^{-i p x_0},\nonumber\\
 &\text{which implies}\nonumber\\
 & \frac{\partial }{\partial t}\left[e^{r t + D_T t  p^2 + D_A t \frac{p^2}{1+a_0^2 p^2}} \Tilde{P}_r(p,t)\right]=re^{r t + D_T t  p^2+ D_A t \frac{p^2}{1+a_0^2 p^2}} e^{-i p x_0}.
\end{align}
Integrating the above, one can obtain 
$$\Tilde{P}_r(p,t)- \Tilde{P}_r(p,0)e^{-r t -D_T t  p^2-D_A t \frac{p^2}{1+a_0^2 p^2}}= r \int_{0}^{t}dt_1 e^{-r (t-t_1) + D_T  p^2 (t-t_1)+ D_A  \frac{p^2}{1+a_0^2 p^2}(t-t_1)} e^{-i p x_0}.$$
Taking $x_0=0,$ and using Eq. (\ref{FP0_fourier}), 
$ \Tilde{P}_0(p,t)=e^{-D_T p^2 t-D_A t \frac{p^2}{1+a_0^2 p^2}},$
 one  can arrive at 
Eq. (\ref{renewal_Fourier}).

 \section{Long-time behavior of Eq. (\ref{P_r_final}) \label{appen3}}
Following the procedure  described in Ref. \cite{goswami2019diffusion},   we rewrite Eq. (\ref{renewal_Fourier})  as 
 \begin{align}
  &\Tilde{P}_r(p,t)=e^{-r t}e^{-D_T p^2 t-D_A t \frac{p^2}{1+a_0^2 p^2}}+r\int_{0}^{t}\, dt'\,e^{-r t'}\,e^{-D_T p^2\,t'-\mu\,t'} \sum_{n=0}^{\infty} (\frac{(\mu\,t')^n}{n!} \left(\frac{1}{1+ a_0^2 p^2}\right)^n\nonumber\\
  &=e^{-r t}\,e^{-D_T p^2 t-D_A t \frac{p^2}{1+a_0^2 p^2}}+r\int_{0}^{t}\, dt'\,e^{-r t'-D_T p^2\,t'-\mu \,t'}+r\int_{0}^{t}\, dt'\,e^{-r t'-D_T p^2\,t'-\mu\,t'}\sum_{n=1}^{\infty}\frac{(\mu\,t')^n}{n!} \left(\frac{1}{1+ a_0^2 p^2}\right)^n\label{approx_FourierP}
 \end{align}
 For large $t$, the first term on the RHS diminishes to zero. So  we only consider the other terms and rewrite  Eq. (\ref{approx_FourierP}) as
 \begin{align}
\Tilde{P}_r(p,t) &\approx  r\int_{0}^{t}\, dt'\,e^{-r t'-D_T p^2\,t'-\mu \,t'}+r\int_{0}^{t}\, dt'\,e^{-r t'-D_T p^2\,t'-\mu\,t'}\sum_{n=0}^{\infty} \frac{(\mu\,t')^{n+1}}{(n+1)!} \left(\frac{1}{1+ a_0^2p^2}\right)^{n+1}\nonumber\\
& \approx  r\int_{0}^{t}\, dt'\,e^{-r t'-D_T p^2\,t'-\mu \,t'}\nonumber\\
 &+r\int_{0}^{t}\, dt'\,e^{-r t'-D_T p^2\,t'-\mu \,t'}\sum_{n=0}^{\infty} \frac{(\mu\,t')^{n+1}}{(n+1)! n!}\left\{\left(-\frac{\partial}{\partial \beta}\right)^n\int_{0}^{\infty}d\phi\,(2 \phi \mu t') e^{-\phi^2\mu  t'(\beta+a_0^2 p^2)}\right\}_{\beta=1}.
 \end{align}
On performing the  inverse Fourier transform of the above and doing the derivative with respect to $\beta$ at $\beta=1,$ we have
\begin{align}
  P_r(x,t)&\approx r\int_{0}^{t}\, dt'\,e^{-r t'-\mu\,t'} \frac{e^{-\frac{x^2}{4 D_T t'}}}{\sqrt{4 \pi D_T t'}}\nonumber\\
  &+2 r\int_{0}^{t}\, dt'\,\int_{0}^{\infty}d\phi\,e^{-r t'-\mu\,t'-\phi^2 \mu t'}\sum_{n=0}^{\infty} \frac{(\mu\,t')^{2(n+1)}\,\phi^{2n+1}}{(n+1)! n!}\frac{e^{-\frac{x^2}{4 t(D_T+D_A \phi^2)}}}{\sqrt{4 \pi t(D_T+D_A \phi^2)}}\nonumber\\
  &\approx r\int_{0}^{t}\, dt'\,e^{-r t'-\mu\,t'} \frac{e^{-\frac{x^2}{4 D_T t'}}}{\sqrt{4 \pi D_T t'}}\nonumber\\
  &+2 r\int_{0}^{t}\,(\mu t') dt'\,\int_{0}^{\infty}d\phi\,e^{-r t'-\mu\,t'-\mu  \phi^2 t'}\,I_1(2\mu  t' \phi)\frac{e^{-\frac{x^2}{4 t(D_T+D_A \phi^2)}}}{\sqrt{4 \pi t(D_T+D_A \phi^2)}}.\label{approx_P}
\end{align}
 In  the second step, the sum over $n$ results in $I_1(z),$ which is the modified Bessel function of the first kind.  Two useful limiting values of $I_1(z)$ are $I_1(z) \approx \frac{z}{2}$ for  $z\ll 1,$ and $I_1(z)\approx \frac{e^{z}}{\sqrt{2\pi z}}$ as $z \gg 1.$

\end{document}